%Paper: hep-th/9502117
%From: minic@scisun.sci.ccny.cuny.edu (DJORDJE MINIC)
%Date: Sat, 18 Feb 95 20:07:48 EST
%Date (revised): Wed, 22 Feb 95 19:25:36 EST

\font\twelvebf=cmbx12
\font\ninerm=cmr9
\nopagenumbers
\magnification =\magstep 1
\overfullrule=0pt
\baselineskip=18pt
\line{\hfil CCNY-HEP 1/95}
\line{\hfil February 1995}
\vskip .8in
\centerline{\twelvebf Remarks on Large N Coherent States}
\vskip .5in
\centerline{\ninerm D. MINIC }
\vskip .1in
\centerline{ Physics Department}
\centerline{City College of the City University of New York}
\centerline{New York, New York 10031.}
\centerline {E-mail: minic@scisun.sci.ccny.cuny.edu}
\centerline{ }
\vskip 1in
\baselineskip=16pt
\centerline{\bf Abstract}
\vskip .1in

Analogs of ordinary Gaussian coherent states on bosonic Fock spaces are
constructed for the case of free Fock spaces, which appear to be natural
mathematical structures suitable for description of large N matrix models.

\vfill\eject

\footline={\hss\tenrm\folio\hss}

\def \12 {{\textstyle {1\over 2}}}

\magnification =\magstep 1
\overfullrule=0pt
\baselineskip=22pt
\pageno=2
%\noindent{\bf 1. Introduction}
\vskip .2in
1. Basic idea:
Gaussian coherent states (for a clear discussion see for example [1]) describe
extremely well ground states of large N theories  with fields in the
fundamental
representation of the symmetry group (so called vector models). On the other
hand
the same Gaussian coherent states fail to describe  ground states of large N
theories with fields in the adjoint representation ( so called matrix models).
(This was clearly emphasized by Yaffe and co-workers [2]). The question arises
if
there is a natural analog to Gaussian coherent states  even for matrix models.
Recall that these states  are inherent to  the structure of the
bosonic Fock spaces. Recently it has become apparent  that another algebraic
structure seems to be natural for the large N matrix models,
the so called free Fock space . This concept naturally appears within the
context of non-commutative probability theory, developed by Voiculescu and
collaborators [3], and it has been used in the
physics literature
 in the analysis of large N matrix models by Haan [4] and  Cvitanovi\'{c} and
co-workers [5] , and more recently by Douglas [6],[7] and
Gopakumar and Gross [8].(Similar ideas have been exploited  in the study of so
called infinite statistics by  Greenberg [9]). It thus seems reasonable to  ask
if
one can find  analogous coherent states related to the structure of
the free Fock space, which would  hopefully play a role similar to ordinary
Gaussian coherent states.
In this short note we examine this question from a rather elementary point of
view. Nevertheless this simple exercise seems to be needed if we are to get
more familiar with the physics contained in the free Fock spaces.
The note is organized as follows: First we review the construction of familiar
Gaussian coherent states  on a bosonic Fock space and then following the same
steps we construct their analogs in the case of a free Fock space. We also
comment on the uniqueness of the construction.

2. Coherent states on a bosonic Fock space: Let us briefly review some
well-known
facts about Gaussian coherent states. Start with the usual bosonic algebra
$[a,a^{\dagger}]=1$,
$a|0>=0$. (The bosonic Fock space being defined as
$|k>: |0>, a^{\dagger}|0>,{1 \over \sqrt{2!}} (a^{\dagger})^2|0>,...,
{1 \over \sqrt{n!}} (a^{\dagger})^n |0>,...$;
$<n|m>=\delta_{n,m}$). Coherent states are then generated by the following
formally unitary
operator (whose form is related to the phase-space structure arising in the
classical limit)
$$
U(z,\bar{z})=\exp(z a^{\dagger} -\bar{z} a),
$$
 or more specifically
$$
|z>= U(z, \bar{z}) |0>.
$$
The states have the following familiar form
$$
|z>=\exp(-{|z|^2 \over 2}) \sum_{n=0}^{\infty} {z^n \over \sqrt{n!}} |n>.
$$
 It follows by construction  that $<z|z>=1$. The property of the
 "resolution of unity" is true as well
$$
{1 \over \pi} \int d^2 z |z><z| =1
$$
 implying the self-reproducing property of the  Gaussian kernel
$$
<z'|z>=\exp(-{|z|^2 \over 2}+\bar{z'} z -{|z'|^2 \over 2})
$$
 or in other words
$$
<z_1|z_2>={1 \over \pi} \int d^2 z <z_1|z><z|z_2>.
$$
The creation and annihilation operators have the following simple actions:
$<z|a^{\dagger}=\bar{z}<z|$ and $a|z>=z|z>$. These relations  follow from the
fact that
$$
0=U(z, \bar{z}) a|0>=U(z, \bar{z}) a U^{-1}(z, \bar{z})U(z, \bar{z})|0>
$$
and
$$
U(z, \bar{z}) a U^{-1}(z, \bar{z})= a-z
$$
by  the use of the Baker-Hausdorff formula.

3. Coherent states on a free Fock space: Let us try to emulate above procedure
in
the  case of the so
called Cuntz algebra [3] $a a^{\dagger}=1$, $a|0>=0$ (The corresponding free
Fock
space being defined as $|k>: |0>, a^{\dagger}|0>, (a^{\dagger})^2
|0>,...,(a^{\dagger})^n |0>,...$;
$<n|m>=\delta_{n,m}$). From completeness $a^{\dagger} a = 1 - |0><0|$ follows
$[a,a^{\dagger}]=|0><0|$.
As in the previous case look at the states
$$
|z>= U(z, \bar{z}) |0>
$$
with
$$
U(z,\bar{z})=\exp(z a^{\dagger} -\bar{z} a).
$$
By expanding the exponential and using the basic  properties  of  the
annihilation
and creation operators, namely $a|n>=|n-1>$ and $a^{\dagger}|n>=|n+1>$,
we are led to
$$
|z>= \sum_{n=0}^{\infty} (n+1) J_{n+1}(2|z|) |z|^{-(n+1)} z^n |n>
$$
where $J_n$ denotes the Bessel function of order n.
We have used the well-known expansion
$$
J_{n}(x)=\sum_{i=0}^{\infty} { {(-1)^{i} x^{2i+n}} \over {2^{2i+n} i!(i+n)!}}.
$$
By construction (and by virtue of the identity
$\sum_{n=0}^{\infty} (n+1)^2 {J_{n+1}(2|z|)}^2=|z|^2$; special case of
Watson's
2.72(1) [10]), $<z|z>=1$. Unfortunately, it turns out that
$$
{1 \over \pi}\int d^2 z <n|z><z|m> = (n+1) \delta_{nm}.
$$
(Here we have made use of $ \int_{0}^{\infty} dr J_{m}^{2} (2r) r^{-1} = {1
\over
2m}$;  special case of Watson's 13.41(2) [10].)

We therefore define a new set of states with the automatic property of the
"resolution of unity"
$$
|z>= \sum_{n=0}^{\infty} \sqrt{n+1} J_{n+1}(2|z|) |z|^{-(n+1)} z^n |n>.
\eqno(1)
$$
The self-reproducing kernel is given by
$$
K(z',z)\equiv <z'|z>=\sum_{n=0}^{\infty} (n+1) {J_{n+1}}(2|z|) {J_{n+1}}(2|z'|)
(|z||z'|)^{-(n+1)} (z \bar{z'})^n . \eqno(2)
$$
Using the multiplication formula for the Bessel functions ( with definitions:
$z=r_1 e^{i\theta_{1}}$ and $z'=r_2 e^{i\theta_{2}}$), to wit
$$
J_{n-k}(r_1) J_{k}(r_2) ={1 \over 2\pi}\int_{0}^{2\pi} d\beta \exp(in\alpha
-ik\beta) J_{n}(r)
$$
(where $r^2= r_1^2 +2r_1 r_2 \cos{\beta} + r_2^2$ and
$r\exp(i\alpha)=r_1+r_2\exp(i\beta)$; Watson's 11.3(3) [10]; for a beautiful
derivation of this formula consult [12]) as well as the integral
representation $
J_{n}(r) = {1
\over 2\pi}\int_{0}^{2\pi}d\phi
\exp(ir\sin{\phi}-in\phi)$ and the recursion formula $2nJ_{n}(r)=
r(J_{n-1}(r)+J_{n+1}(r)$) we get ($\theta\equiv \theta_{1} -\theta_{2}$)
$$
K(z',z)={e^{-i\theta} \over {8\pi^{2}r_1 r_2}}\int_{0}^{2\pi} d\beta
\int_{0}^{2\pi} d\phi
r \cos{\phi} \exp(ir\sin{\phi}) \sum_{n=1}^{\infty}
e^{in(2\alpha - \beta +\theta - 2\phi)}. \eqno(3)
$$
If $z=z'$ ($r_1=r_2=R$, $2\alpha=\beta$ and $\theta=0$) we get,
after doing the $\beta$-integral first and then the $\phi$-integral,
$$
F(R)\equiv <z|z>= \sum_{k=0}^{\infty}{{(-1)^{k}R^{2k}(2k)!} \over {[(k+1)!]^{2}
[k!]^{2}}}. \eqno(4)
$$
It is apparent  that the behavior  of  $<z|z>$ around $R=0$ is regular. The sum
is
evidently convergent and by construction $<z|z>$ is positive definite (as it
should be if  $|z>'s$  are to form a Hilbert space, which they do by
construction).

 We quote a number of other representations of $K(z',z) $ and $F(R)$. Another
integral representation implied by the multiplication formula and
$J_{-n}=(-1)^{n} J_{n}$ is
$$
K(z',z)=-{1 \over {2\pi r_1 r_2}} \int_{0}^{2\pi} d\beta
{{e^{i \beta} J_{0}(2r)} \over {[1+e^{i \theta} e^{i \beta}]^{2}}}.  \eqno(5)
$$
Apparently the most explicit expression for $K(z',z)$ is given by the following
double-sum representation
$$
K(z',z)={e^{-i\theta} \over {r_1 r_2}}\sum_{n=1}^{\infty} \sum_{k=1}^{\infty}
e^{i n\theta} (-1)^{k} {{n i^{n}} \over {k!(k+n)!}} P_{k}^{n}(r_{1}^{2}
+r_{2}^{2})  \eqno(6)
$$
where $P_{k}^{n}(x)$ denotes the associated Legandre functions.
(Here we have made use of the following well-known integral representation
of  $P_{k}^{n}(x)$; Whittaker and Watson's 15.61 [11]),
$$
P_{k}^{n}(\cos{\gamma})={{i^{n}(n+k)!} \over {2\pi k!}} \int_{0}^{2\pi}
d\varphi
(\cos{\gamma} + i \sin{\gamma} \cos{\varphi})^{k} e^{i n \varphi}.
$$
The above  double sum reduces to the single sum formula for $<z|z>$ which
of course
 follows from the integral representation
$$
F(R)=-{1 \over {2\pi R^2}} \int_{0}^{2\pi} d\beta
{{ J_{0}(4 R \cos{{\beta \over 2}})} \over {4 [\cos{{\beta \over 2}}]^{2}}},
\eqno(7)
$$
 which is a special case of (5).
By utilizing the following remarkable property of Bessel functions; Watson's
5.4(5) [10]
$$
J_{\nu}(z)J_{\mu}(z)=\sum_{m=0}^{\infty}{{(-1)^{m}({z \over 2})^{\mu + \nu +2m}
\Gamma(\mu + \nu + 2m +1)} \over {m! \Gamma(\mu + m+ 1) \Gamma(\nu + m +1)
\Gamma(\mu + \nu + m + 1)}}
$$
we arrive at
$$
R^{2}{dF(R) \over dR} + R F(R) = J_{0}(2R)J_{1}(2R)
$$
implying yet another integral representation
$$
F(R)={1 \over R}(\int{1 \over R}J_{0}(2R)J_{1}(2R) dR +c)  \eqno(8)
$$
c being a constant of integration such that $F(0)=1$.

Any of the  above  expressions for $K(z',z)$ satisfy  by construction
the self-reproducing property
$$
K(z_1,z_2)= {1 \over \pi} \int d^2 z K(z_1,z)K(z,z_2).
$$

It is also possible to find actions of the operators $a$ and $a^{\dagger}$
but the expressions are a bit unwieldy and we do not state them explicitly.
One fact is evident though, the new coherent states are not the eigenstates
of the annihilation operator $a$.

4.Comments on the uniqueness of new coherent states: One could ask how robust
our construction is. In particular one could ask if one could get more
manageable expressions if one examined the eigenstates of the annihilation
operator $a$ as appropriate coherent states. This question can of course be
easily
answered. Let us  look at $|z>$ such that $a|z> = z|z>$. By expanding
$$
|z>=\sum_{n=0}^{\infty} c_{n}  |n>
$$
and acting with  the operator $a$  on the left we obtain
$$
|z>=\sum_{n=0}^{\infty} z^{n} |n>.
$$
By examining $<z'|z>$ we see that for the reasons of convergence we have  to
constrain
$|z|$ inside the unit circle. Again, the property of the "resolution of unity"
is not
satisfied. So we redefine the states so that this condition is automatically
fulfilled (the new states will thus seize to be eigenstates of $a$)
$$
|z>=\sum_{n=0}^{\infty} \sqrt{n+1} z^{n} |n>. \eqno(9)
$$
The self-reproducing kernel is now
$$
<z'|z> = {1 \over (1- z \bar{z'})^{2}} \eqno(10)
$$
and the normalization condition reads
$$
<z|z> = {1 \over (1- z \bar{z})^{2}}.
$$
Unfortunately this construction, unlike the previous one,
is valid only for $|z| <1$.

The more important question is: why have we at all decided to use the
exponential
representation for the unitary operator $U(z,\bar{z})$, namely
$U(z,\bar{z}) = \exp(z a^{\dagger} - \bar{z} a)$ (which indeed is the
appropriate
one for the Gaussian coherent states)? We do not have a convincing  answer to
this question. We have examined, by trial and error,
a number of different representations for $U(z,\bar{z})$  and only in the case
of
the familiar exponential one have we found an interesting result. It is quite
possible that there is a more natural form for $U(z,\bar{z})$ which would be
more suitable for
coherent states on free Fock spaces.

5. Conclusion: Analogs of familiar Gaussian coherent states are constructed
for the case when the annihilation and creation operators satisfy  the Cuntz
algebra. The result turns out to be more  complicated, but the essential
features
of ordinary Gaussian coherent states are preserved. By doing this fairly simple
exercise some familiarity with unusual free Fock spaces is gained, albeit no
real
insight into relevant physical questions concerning large N matrix models. It
would therefore be extremely interesting to see if these  new states have any
practical physical applications.

\vskip.1in
Numerous helpful discussions with V.P.Nair are more than greatfully
acknowledged.
Many warm thanks to Mike Douglas  for providing inspiration and generous share
of
insights. Last but not least, I thank Bunji Sakita for support and
encouragement.
This work was supported in part by the NSF grant PHY 90-20495 and by the
Professional Staff Board of Higher Education of the City University of New York
under grant no. 6-63351.

\vskip.1in
{\bf References}
\item{1.} J.R.Klauder and E.C.G.Sudarshan, {\sl Fundamentals of Quantum
Optics},
Benjamin, 1968.
\item{2.} L.G.Yaffe, Rev.Mod.Phys. 54 (1982) 407; F.Brown and L.G.Yaffe,
Nucl.Phys. B271 (1986) 267; T.A.Dickens, U.J.Lindqwister, W.R.Somsky and
L.G.Yaffe, Nucl.Phys. B309 (1988) 1.
\item{3.} D.V.Voiculescu, K.J.Dykema and A.Nica, {\sl Free Random Variables},
AMS, Providence 1992.
\item{4.} O.Haan, Z.Phys C6 (1980) 345; see also M.B.Halpern and C.Schwarz,
Phys.Rev. D24 (1981) 2146 and A.Jevicki and H.Levine, Ann.Phys. 136 (1981) 113.
\item{5.} P.Cvitanovi\'{c}, Phys.Lett. 99B (1981); P.Cvitanovi\'{c},
P.G.Lauwers and
P.N.Scharbach, Nucl.Phys. B203 (1982) 385.
\item{6.} M.R.Douglas, "Large N Gauge theory - Expansions and Transitions", to
appear in the proceedings of the 1994 ICTP Spring School, hep-th/9409098.
\item{7.} M.R.Douglas, "Stochastic Master Fields", RU-94-81, hep-th/9411025;
M.R.Douglas and M.Li, "Free Variables and the Two Matrix Model", BROWN-HET-976,
RU-94-89, hep-th/9412203.
\item{8.} R.Gopakumar and D.J.Gross, "Mastering the Master Field", PUPT-1520,
hep-th/9411021.
\item{9.} O.Greenberg, Phys.Rev.Lett. 64 (1990) 705.
\item{10.} G.N.Watson, {\sl A Treatise on the Theory of Bessel Functions}, 2nd
edition, Cambridge University Press, 1944.
\item{11.} E.T.Whittaker and G.N.Watson, {\sl A Course in Modern Analysis}, 4th
edition, Cambridge University Press, 1927.
\item{12.} N.J.Vilenkin, {\sl Special Functions and the Theory of Group
Representation}, AMS, Providence, 1968.

\end